\documentclass[transmag]{IEEEtran}
\usepackage{latexsym}
\usepackage{graphicx}
\usepackage{amsfonts,amssymb,amsmath}
\usepackage{hyperref}

\usepackage{multirow}

\usepackage{color}
\usepackage[caption=false,font=footnotesize]{subfig}

\newcommand{\varLatent}{\mathbf{y}}
\newcommand{\varquantLatent}{\hat{\mathbf{y}}}
\newcommand{\varnoisyLatent}{\tilde{\mathbf{y}}}
\newcommand{\varInp}{\mathbf{x}}
\newcommand{\varRec}{\hat{\mathbf{x}}}
\newcommand{\varQuant}{\mathrm{Q}}
\newcommand{\varEntCode}{\mathrm{EC}}
\newcommand{\varEntDecode}{\mathrm{ED}}
\newcommand{\Expect}{{\mathsf{E}}}

\newcommand{\modes}{\mathbf{m}}
\newcommand{\logar}{\mathrm{log}} 

\newcommand{\btheta}{\mathbf{\theta}}
\newcommand{\bphi}{\mathbf{\phi}}

\def\BibTeX{{\rm B\kern-.05em{\sc i\kern-.025em b}\kern-.08em T\kern-.1667em\lower.7ex\hbox{E}\kern-.125emX}}
\begin{document}

\title{Transform Network Architectures for Deep Learning based End-to-End Image/Video Coding in Subsampled Color Spaces}

\author{Hilmi E. Egilmez, 
Ankitesh K. Singh, Muhammed Coban, Marta Karczewicz \\ Yinhao Zhu, Yang Yang, Amir Said, 
Taco S. Cohen 
\thanks{Earlier versions of this work were presented in \cite{Egilmez:21:E2E_JVET_U0079,Singh:21:E2E_JVET_U0080} as JVET standard contributions.}
\thanks{H.~E.~Egilmez, A.~K.~Singh, M.~Coban, M.~Karczewicz, Y.~Zhu, Y.~Yang and A.~Said are with 
Qualcomm Technologies, Inc., San Diego, CA 92121 USA (e-mail: \{hegilmez, ankitesh, mcoban, martak, yinhaoz, yyangy, asaid\}@qti.qualcomm.com).}
\thanks{ T.~S.~Cohen is with Qualcomm Technologies Netherlands B.V., Amsterdam, 1098 XH Netherlands (e-mail: tacos@qti.qualcomm.com).}
}

\IEEEtitleabstractindextext{\begin{abstract}
Most of the existing deep learning based end-to-end image/video coding (DLEC) architectures are designed for non-subsampled RGB color format. However, in order to achieve a superior coding performance, many state-of-the-art block-based compression standards such as High Efficiency Video Coding (HEVC/H.265) and Versatile Video Coding (VVC/H.266) are designed primarily for YUV 4:2:0 format, where U and V components are subsampled by considering the human visual system. This paper investigates various DLEC designs to support YUV 4:2:0 format by comparing their performance against the main profiles of HEVC and VVC standards under a common evaluation framework. Moreover, a new transform network architecture is proposed to improve the efficiency of coding  YUV 4:2:0 data. The experimental results on YUV 4:2:0 datasets show that the proposed architecture significantly outperforms naive extensions of existing architectures designed for RGB format and achieves about 10\% average BD-rate improvement over the intra-frame coding in HEVC.  
\end{abstract}

\begin{IEEEkeywords}
Deep learning, neural networks, transform network, data compression, image coding, video coding, color spaces, YUV, RGB.
\end{IEEEkeywords}

}

\maketitle

\section{INTRODUCTION}
\begin{figure}[!t]
\centering
    \subfloat[Building blocks used in \emph{training} with noise 
    ($p_{\text{noise}}$) added to generate noisy latent coefficients ($\varnoisyLatent$) for approximating non-differentiable quantization impact in the bottleneck.\label{fig:vae}]{\includegraphics[width=0.95\columnwidth]{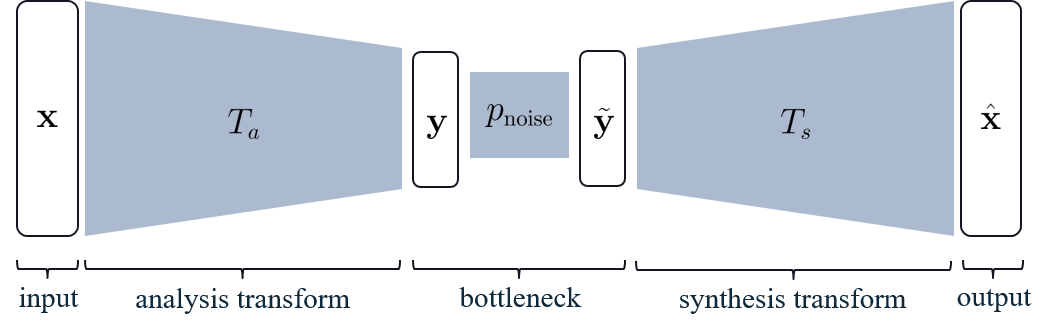}} \\
    \subfloat[Building blocks used in \emph{compression/{testing}} consisting of quantization ($\varQuant$), entropy coding ($\varEntCode$), and entropy decoding ($\varEntDecode$) in the bottleneck yielding quantized latent coefficients ($\varquantLatent$). \label{fig:vae_entropy}]{\includegraphics[width=0.95\columnwidth]{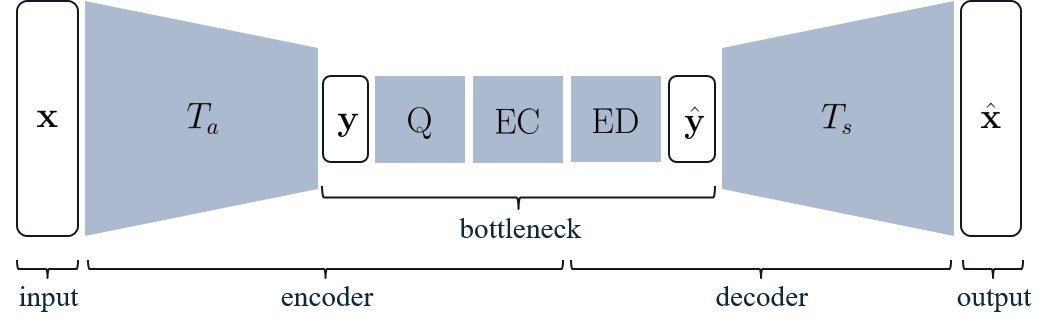}}
\caption{Illustration of basic VAE-oriented DLEC architectures consisting of multiple convolutional layers denoted by $T_a$ and $T_s$, where the analysis (forward) transform generates latent coefficients ($\mathbf{y}$) from a given input ($\varInp$), and the synthesis (inverse) transform $T_s$ yields the reconstruction $\varRec$. Since quantization {($\varQuant$) used in actual compression} is non-differentiable and prevents end-to-end training, {typically} uniform noise {($p_{\text{noise}}$)} is added as a proxy to quantization {when training}.}
\label{fig:vae_overall}
\end{figure}

\IEEEPARstart{R}{ecent} research studies on deep learning based end-to-end image or intra-frame video coding (DLEC) have shown that competitive compression performances can be achieved as compared to existing block-based compression standards in coding RGB (red, green and blue) sources. 
These DLEC approaches typically employ an end-to-end 
variational autoencoder (VAE) \cite{Kingma:2014:VAE} architecture, 
illustrated in Fig.~\ref{fig:vae}, 
where neural networks are trained on sources represented in RGB format based on a rate-distortion (RD) criterion. Those trained models are evaluated by compressing the RGB data as shown in Fig.~\ref{fig:vae_entropy}, where the transformed input data is quantized and entropy coded at the encoder side, then the bitstream is decoded and the resulting quantized coefficients are inverse transformed to reconstruct the RGB data. 
Yet, most practical image and video compression systems operate on luminance (luma) and chrominance (chroma) components, represented in a YUV format. 

\begin{figure}[!t]
\centering
    \subfloat[RGB/YUV 4:4:4\label{fig:yuv444}]{\includegraphics[width=0.4\columnwidth]{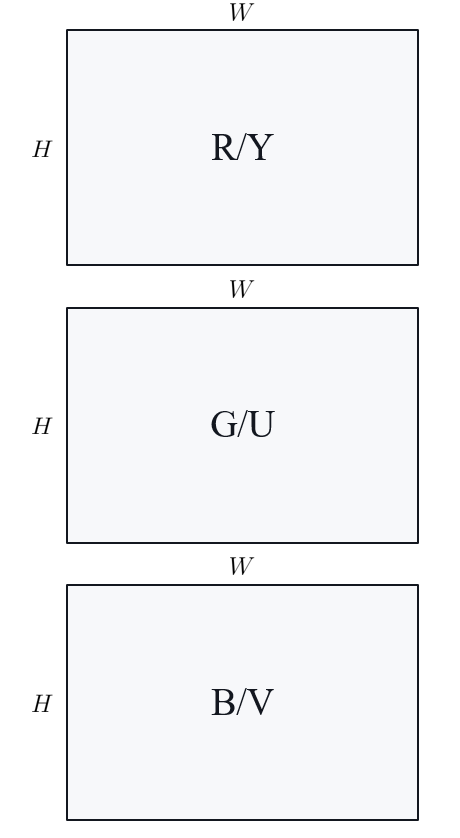}} \quad 
    \subfloat[YUV 4:2:0\label{fig:yuv420}]{\includegraphics[width=0.4\columnwidth]{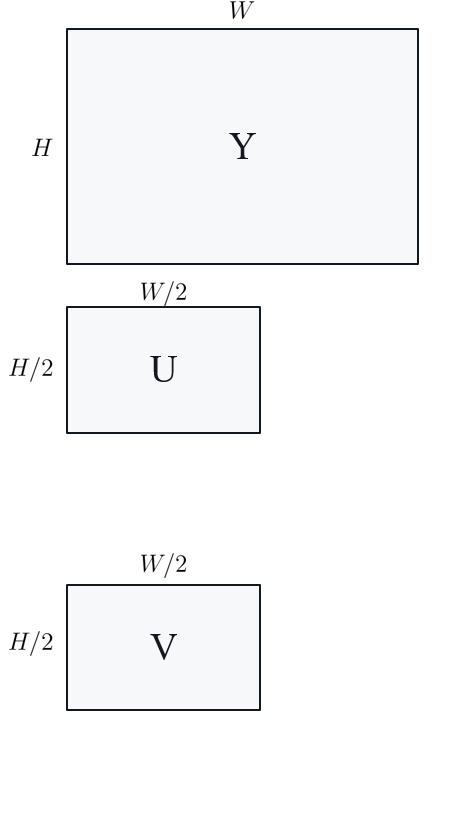}} 
\caption{Illustration of (a) non-subsampled RGB/YUV 4:4:4 and (b) YUV 4:2:0 with chroma subsampling reducing chroma resolution by four times for a frame with $W \times H$ resolution.}    
\label{fig:yuv_formats}
\end{figure}

\begin{figure}[!t]
\centering
    {\includegraphics[width=0.9\columnwidth]{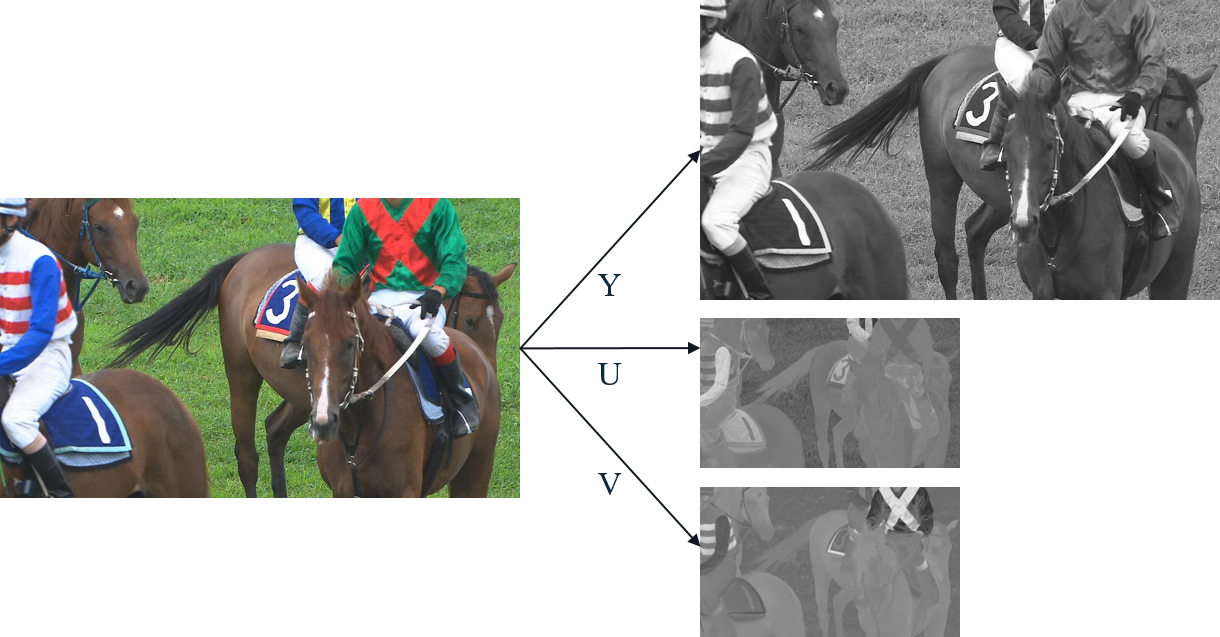}}
\caption{An example of YUV 4:2:0 decomposition on a frame in \emph{Racehorses} video sequence \cite{Bossen:18:ctc}, where Y component is in fact the grayscale version of the original colored frame, and U and V components capture color information. }
\label{fig:yuv_decomposition}
\end{figure}

\begin{figure}[!t]
\centering
    {\includegraphics[width=\columnwidth]{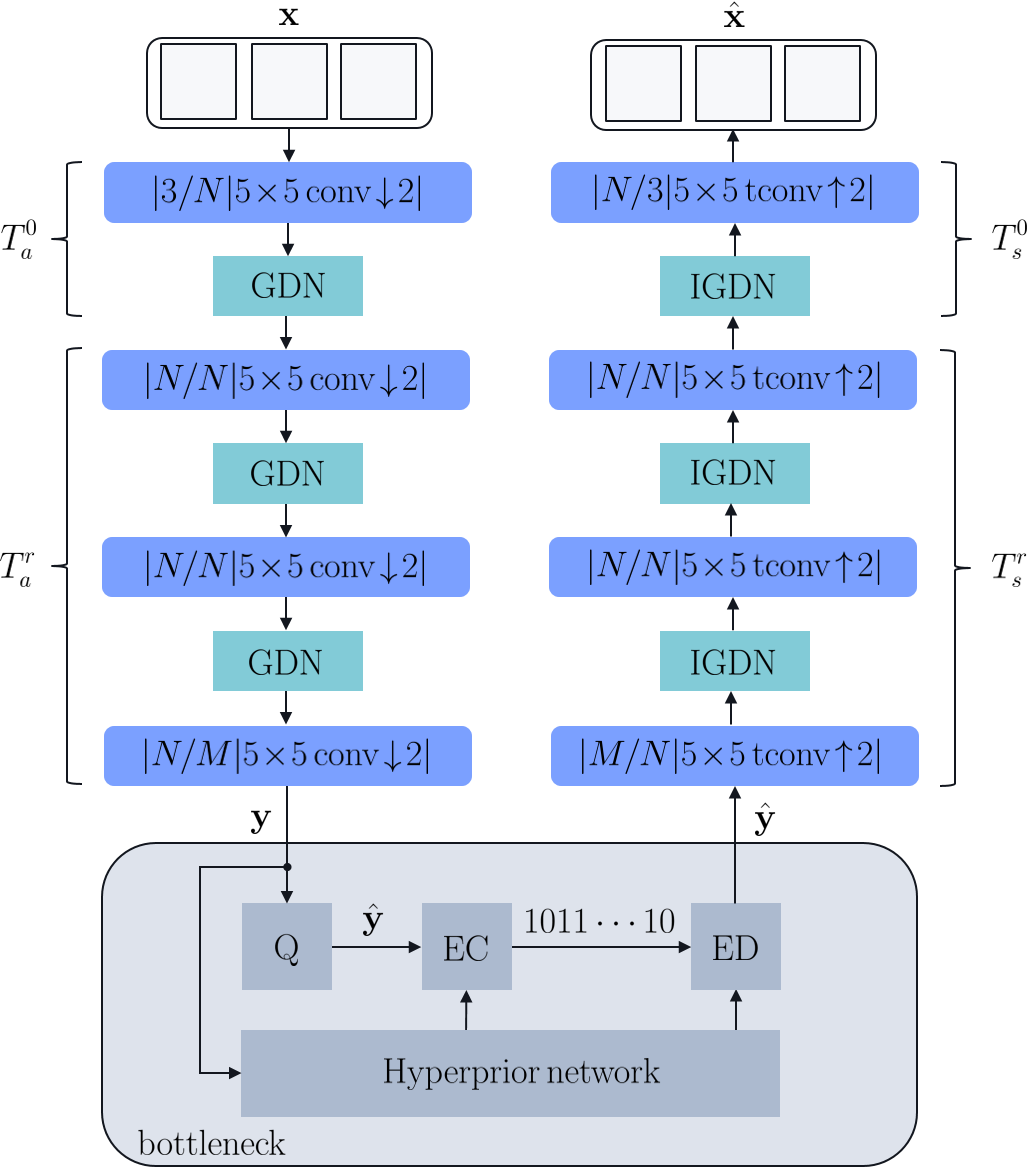}} 
\caption{The transform network architecture used in \cite{Balle:2018:ICLR:ScaleHyper,MinnenBT:2018:NeurIPS:MSHyper,Minnen:2020:ICIP:Charm}, which focus on different bottleneck and hyperprior network designs with the same transform network (i.e., pair of $T_a$ and $T_s$). Moreover, $T_a^0$, $T_a^r$ $T_s^0$ and $T_s^r$ denote the transform sub-networks that will be used in Section \ref{sec:dnn_archs}. Note that they satisfy $\varLatent=T_a(\varInp)=T_a^r(T_a^0(\varInp))$ and $\varRec = T_s(\varquantLatent)=T_s^0(T_s^r(\varquantLatent))$.}    
\label{fig:google_transfrom_network}
\end{figure}

Among various color formats, YUV 4:2:0 is predominantly adopted as the basic input-output format in many state-of-the-art compression standards, which include the main profiles\footnote{The main profile of a video coding standard defines the most basic set of instructions that needs be implemented to support the standard in both hardware and software. In case of HEVC and VVC, their main profiles only support YUV 4:2:0 format and the other formats such as RGB and YUV 4:4:4 are considered as extensions and supported optionally.} of High Efficiency Video Coding (HEVC) \cite{Sullivan:12:hevc} and Versatile Video Coding (VVC) \cite{Bross:20:vvc10}. 
As shown in Fig.~\ref{fig:yuv_formats}, unlike non-subsampled RGB and YUV 4:4:4 formats, YUV 4:2:0 consists of subsampled chroma components (U and V), 
while the luma component (Y) is not subsampled and retained at the same resolution. 
The main reasons for using YUV 4:2:0 format for image/video coding can be summarized as follows:
\begin{itemize}
\item \emph{Human Visual System (HVS)}: For HVS, the luma component is much more important than chroma components, since human eye is far less sensitive to color details (captured by the chroma components) than to brightness details (in the luma component) \cite{Winkler:2001:HVS,Pearlman:2011:DigSigComp,Tekalp:2015:DVP}. To take advantage of this HVS behavior, U and V components in YUV 4:2:0 are subsampled without degrading the perceptual quality. An example YUV decomposition in Fig.~\ref{fig:yuv_decomposition} perceptually demonstrates that Y component carries more salient details than U and V components do. 
\item \emph{Compression benefit}: In video compression, chroma subsampling in YUV 4:2:0 greatly reduces the total bitrate, because chroma components at a lower resolution are generally coded in fewer bits. Based on our experiments with the HEVC standard, coding in YUV 4:2:0 format yields about a 20\% less bitrate on average as compared to coding the same video content in non-subsampled YUV 4:4:4. 
\item \emph{Complexity reduction}: As compared to YUV 4:4:4 format, supporting YUV 4:2:0 on hardware and software implementations is less complex in terms of memory, number of computations and run-time due to processing chroma components at a resolution that is four times smaller than their original resolution.  
\end{itemize}
However, in the literature, there is very little or no work on DLEC designs specialized for YUV sources. Although existing architectures designed for coding RGB data \cite{Balle:2017:ICLR:ImageComp, Balle:2018:ICLR:ScaleHyper,MinnenBT:2018:NeurIPS:MSHyper, chen2019neural,Minnen:2020:ICIP:Charm, Cheng_2020_CVPR} (such as the one shown in Fig.~\ref{fig:google_transfrom_network}) can be employed to support non-subsampled YUV 4:4:4 format by simply retraining network parameters on a YUV 4:4:4 dataset, effective solutions for chroma subsampled formats, such as YUV 4:2:0, are non-trivial and require new neural network architectures. 
{Moreover, the work in \cite{Ma:2020:E2E_wavelet} introduces a lifting-oriented wavelet scheme using neural networks for prediction and update steps \cite{Sweldens:1996:Liftingscheme}, where the network parameters are trained on luma data, and the trained transformation is applied for coding each color component separately. While the authors of \cite{Ma:2020:E2E_wavelet} do not consider joint coding of YUV data, in this paper we focus on transformation network designs specialized for joint coding of subsampled YUV data.}

This paper investigates various \emph{transform network} solutions 
to support subsampled color formats in VAE-oriented DLEC architectures, where the main goal is to design a pair of analysis and synthesis transform networks, denoted by $T_a$ and $T_s$ in Fig.~\ref{fig:vae_overall}.
Although the focus of this paper is on designs that directly work with YUV 4:2:0 format, the methodology presented in this paper can be simply extended for other types of subsampled color formats (e.g., YUV 4:2:2). 
In order to support YUV 4:2:0 format on DLEC architectures, 
we introduce two distinct approaches. 
The first approach aims to modify the structure of input and output channels (with a predefined method of splitting) so that the dimensions of each channel are equal. Then, the modified input-output channels are utilized in existing DLEC designs for non-subsampled data without any major changes in the architecture. On the other hand, the second approach proposes to build {new transform network architectures} that operate directly on YUV 4:2:0 data without applying any modifications/processing on input or output channels. 
Since pre-processing on raw input data (e.g., channel splitting and color format conversion) can potentially lead to undesired distortion (in YUV 4:2:0 color space) as well as perceptual artifacts that cannot be efficiently recovered at the decoder side, it is ideal (for achieving a better coding performance) to build a network architecture specialized for YUV 4:2:0 format, whose parameters are trained on a YUV 4:2:0 dataset. Thus, the network has complete and trainable control over luma and chroma fidelity in YUV 4:2:0 color space considering HVS. 
In our experimental results, solutions based on these two approaches are comprehensively evaluated and compared against HEVC and VVC under a common evaluation framework used in HEVC and VVC standardization activities \cite{Bossen:13:hevc_ctc,Bossen:18:ctc}.

To the best of our knowledge, this is the first paper in the literature that proposes a novel DLEC architecture specialized for YUV 4:2:0 format. The proposed architecture allows us to balance the coding gains between Y, U and V components considering HVS, which cannot be achieved by the existing architectures designed for RGB. Moreover, noting that HEVC and VVC standards are explicitly designed and optimized for YUV 4:2:0 format, the RGB data compression results in existing DLEC studies do not truthfully reflect the actual performance of state-of-the-art codecs\footnote{As noted in \cite{Pearlman:2011:DigSigComp} (Section 13.2), YUV representation effectively decorrelates the three components, hence provides a more compact encoding. Based on our experiments using  HEVC, the RGB coding is more than 50\% less efficient than coding the same content in YUV 4:2:0 in terms of bitrate.}. With our proposed solutions supporting YUV 4:2:0 format, a fair comparison against the main profiles of HEVC and VVC is presented.

In the remainder of this paper, Section \ref{section:prelim_related_work} discusses related work in the literature and some preliminaries on rate-distortion optimization. The introduced DLEC solutions are presented in Section \ref{sec:dnn_archs}, and corresponding experimental results are reported and analyzed in Section \ref{sec:results}. Concluding remarks based on empirical results are drawn in Section \ref{sec:conclusion}.

\section{Preliminaries and Related Work}
\label{section:prelim_related_work}
\subsection{Rate-distortion optimized coding}
In traditional block-based image and video coding systems, 
an ideal encoder aims to find the best configuration of coding modes/parameters 
(defined by the compression standard) 
that minimizes the rate-distortion (RD) cost \cite{Ortega:98:rdm}:
\begin{equation}
\label{eqn:rd_cost_general}
J_\lambda(\modes) = D(\modes) + \lambda R(\modes) 
\end{equation}
where the multiplier $\lambda$ steers the trade-off between bitrate ($R$) and distortion ($D$), and the choice of $\lambda$ is typically directly proportional to the level of quantization, set by quantization parameter (QP). 
For a given $\lambda$, the encoder searches over different mode configurations ($\modes$) to find the best configuration ($\modes^*$) that minimizes (\ref{eqn:rd_cost_general}).

In VAE-oriented DLEC approaches (Fig.~\ref{fig:vae_overall}), the analysis transform $T_a(\varInp;\bphi)$ {with parameters $\bphi$} is represented by an inference model following the conditional distribution $p_{\bphi}(\varnoisyLatent|\varInp)$
and the synthesis transform $T_s(\varnoisyLatent;\btheta)$ {with parameters $\btheta$} is {probabilistically characterized} by a generative model $p_\theta(\varInp|\varnoisyLatent)$ 
incorporating a prior $p_\theta(\varnoisyLatent)$ on latent variables \cite{Balle:2017:ICLR:ImageComp}. 
In this construction, given the observations from random variable $\varInp$ and the generative model $p_\btheta(\varInp|\varnoisyLatent)$, 
the variational Bayesian inference deals with approximating the intractable posterior 
$p_{\btheta}(\varnoisyLatent|\varInp)$
with a parametric distribution $q_{\bphi}(\varnoisyLatent|\varInp)$ by minimizing the Kullback-Leibler (KL) divergence between {the distributions} $q_\bphi$ and $p_\btheta$: 
\begin{equation}
\label{eqn:KL_divergence}
\mathrm{KL}[q_\bphi\,||\,p_\btheta] = 
\Expect_{\varnoisyLatent \sim q_\bphi} [\logar\,q_\bphi(\varnoisyLatent|\varInp)] -   \Expect_{\varnoisyLatent \sim q_\bphi} [\logar\,p_\btheta(\varnoisyLatent|\varInp)] 
\end{equation}
where the first term ($\Expect_{\varnoisyLatent \sim q_\bphi} [\logar\,q_\bphi(\varnoisyLatent|\varInp)]$) 
is constant by construction \cite{Balle:2017:ICLR:ImageComp}, so the  minimization of (\ref{eqn:KL_divergence}) is equivalent to minimizing the following loss function\footnote{Note that minimizing (\ref{eqn:rd_cost_neural}) is also equivalent to maximizing the evidence lower bound (ELBO) used in variational Bayesian methods \cite{Kingma:2014:VAE}.} used in VAE training:
\begin{equation}
\label{eqn:rd_cost_neural}
L_\beta(\bphi,\btheta)= 
\underbrace{\Expect_{\varnoisyLatent \sim q_\bphi} [-\logar\,p_\btheta(\varnoisyLatent)] }_{R(\bphi,\btheta)} + \beta 
\underbrace{\Expect_{\varnoisyLatent \sim q_\bphi} [-\logar\,p_\btheta(\varInp|\varnoisyLatent)]}_{D(\bphi,\btheta)}
\end{equation}
which is analogous to the RD cost in (\ref{eqn:rd_cost_general}), where the rate $R(\bphi,\btheta)$ in (\ref{eqn:rd_cost_neural}) is equal to the continuous Shannon entropy of latent variables, and the general form of  distortion $D(\bphi,\btheta)$ boils down to commonly used mean-square error (MSE)\footnote{In traditional block-based codecs, MSE is often used as the distortion term in (\ref{eqn:rd_cost_general}).} if the underlying distributions are assumed to be multivariate Gaussian. 
As $\lambda$ in (\ref{eqn:rd_cost_general}), $\beta$ is also chosen as a parameter (i.e., being part of the generative model) to adjust the trade-off between rate and distortion. 
Note that during training the impact of non-differentiable quantization and exact bitrate calculation can only be approximated. Specifically, in (\ref{eqn:rd_cost_neural}), continuous Shannon entropy is used as a proxy for the actual bitrate, and in \cite{Balle:2017:ICLR:ImageComp} a uniform distribution is incorporated in the inference model to better reflect quantization. 
On the other hand, as depicted in Fig.~\ref{fig:vae_overall}, trained networks can be tested and evaluated by entropy coding the quantized (i.e., scaled and rounded) coefficients $\varquantLatent$ to create actual bitstreams.  

In a nutshell, traditional block-based codecs perform RD-optimization by combinatorially searching for the best mode configuration among the set of candidates defined by a compression standard, while DLEC approaches aim to train RD-optimized networks and parameters on an end-to-end basis.

\subsection{Related work on DLEC approaches}
The related research on DLEC approaches can be mainly categorized into two lines of work, about (i) transform network architectures and (ii)  latent variable modeling and hyperprior network designs. 
In DLEC approaches, transform network designs typically follow an autoencoder architecture (Fig.~\ref{fig:vae_overall}) consisting of convolutional layers with downsampling at the encoder and upsampling at the decoder, which are interleaved with nonlinear activations. Fig.~\ref{fig:google_transfrom_network} explicitly demonstrates the state-of-the-art transform network architecture used in \cite{Balle:2018:ICLR:ScaleHyper,MinnenBT:2018:NeurIPS:MSHyper,Minnen:2020:ICIP:Charm}, where the following notation,
\begin{equation}
|C_{\text{in}}/C_{\text{out}}|K\!\times\!K\,\mathrm{conv}\!\downarrow\!2|,\notag
\end{equation} 
specifically denotes two-dimensional $K\!\times\!K$ convolutions with downsampling by 2 having $C_{\text{in}}$ input and $C_{\text{out}}$ output channels. The corresponding transposed convolutions\footnote{Transposed convolution (i.e., $\mathrm{tconv}$) is also known as deconvolution.} at the decoder is expressed as $|C_{\text{in}}/C_{\text{out}}|K\!\times\!K\,\mathrm{tconv}\!\uparrow\!2|$ with $\uparrow\!2$ standing for upsampling by 2. As for nonlinear operations, the analysis transform uses generalized divisive normalization (GDN)~\cite{Balle:2016:GDNfirst} while its synthesis counterpart applies inverted GDN (IGDN) between the convolutions. It has been shown in~\cite{Balle:2016:GDNfirst,Balle:2018:GDN} that GDNs can considerably improve the RD performance 
by their cross-channel normalization capability and performs better than some commonly used activation operators including rectified linear unit (ReLU) and leaky ReLU. 
Among our transform network designs introduced in Section \ref{sec:dnn_archs}, we propose a solution using cross-channel $1\times1$ convolutions together with parametric ReLU (PReLU) activations as alternatives to GDNs. Our experimental results in Section \ref{sec:results} will show that the proposed solution can outperform the variants using GDNs.  

In order to further improve the RD performance, another VAE-oriented network, called \emph{hyperprior network} as shown in Fig.~\ref{fig:google_transfrom_network}, is employed to effectively learn prior model parameters used as input to the entropy coding\footnote{Since hyperprior networks are VAEs, they have  their own bottlenecks (see in Fig.~\ref{fig:vae_overall}) where hyperprior latent variables are quantized and entropy coded.}. While the initial work in \cite{Balle:2017:ICLR:ImageComp} applies a fully factorized prior model that cannot exploit the statistical dependencies between latent variables, later studies include autoregressive models that can capture spatial \cite{Balle:2018:ICLR:ScaleHyper,MinnenBT:2018:NeurIPS:MSHyper} and cross-channel correlations \cite{Minnen:2020:ICIP:Charm} among latent variables. Since our paper focuses on transform network designs, the solutions introduce in the next section can be used together with any hyperprior model\footnote{The reader is referred to \cite{Balle:2020:NTC,Balle:2018:ICLR:ScaleHyper,MinnenBT:2018:NeurIPS:MSHyper,Minnen:2020:ICIP:Charm} for further details on hyperprior networks and latent variable modeling.}. In our experiments, one of the best performing mean-scale hyperprior network \cite{MinnenBT:2018:NeurIPS:MSHyper} is used  as our entropy model in the bottleneck. 

\section{Transform Network Architectures for Image/Video Coding}
\label{sec:dnn_archs}
This section introduces two classes of solutions to support YUV 4:2:0 format. 
The first class of solutions are based on \emph{input-output channel alignment}, which aim to support YUV 4:2:0 without introducing any major changes to the existing network architectures in the literature \cite{Balle:2017:ICLR:ImageComp, Balle:2018:ICLR:ScaleHyper,MinnenBT:2018:NeurIPS:MSHyper, chen2019neural,Minnen:2020:ICIP:Charm,Cheng_2020_CVPR}. 
On the other hand, second class of solutions proposes a new \emph{transform network architecture} where the main goal is to compress YUV 4:2:0 input data more efficiently. In this paper, we introduce three variants of the proposed new architecture.

\subsection{Naive Solutions via Input-Output Channel Alignment}
\label{subsec:naive_alignment}
DLEC architectures designed for RGB sources 
can be accommodated to support YUV 4:2:0 format by  
\emph{aligning channel dimensions}, which essentially involves: 
\begin{itemize}
\item splitting the input/output data so that dimensions of each channel are equal, and 
\item changing the number of input/output channels to the network, accordingly. 
\end{itemize}
Note that this approach is rather straightforward and does not require any fundamental changes in the network architecture.

\begin{figure}[!t]
\centering
    \subfloat[Network used for coding Y component\label{fig:separate_ch_y}]{\includegraphics[width=0.8\columnwidth]{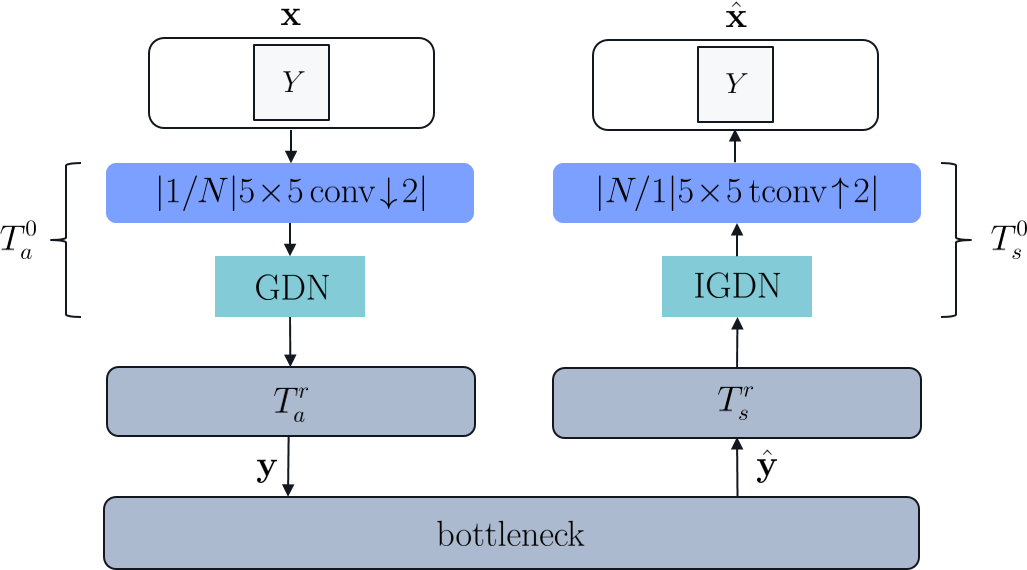}} 
    \\ 
    \subfloat[Network used for coding U and V components\label{fig:separate_ch_uv}]{\includegraphics[width=0.8\columnwidth]{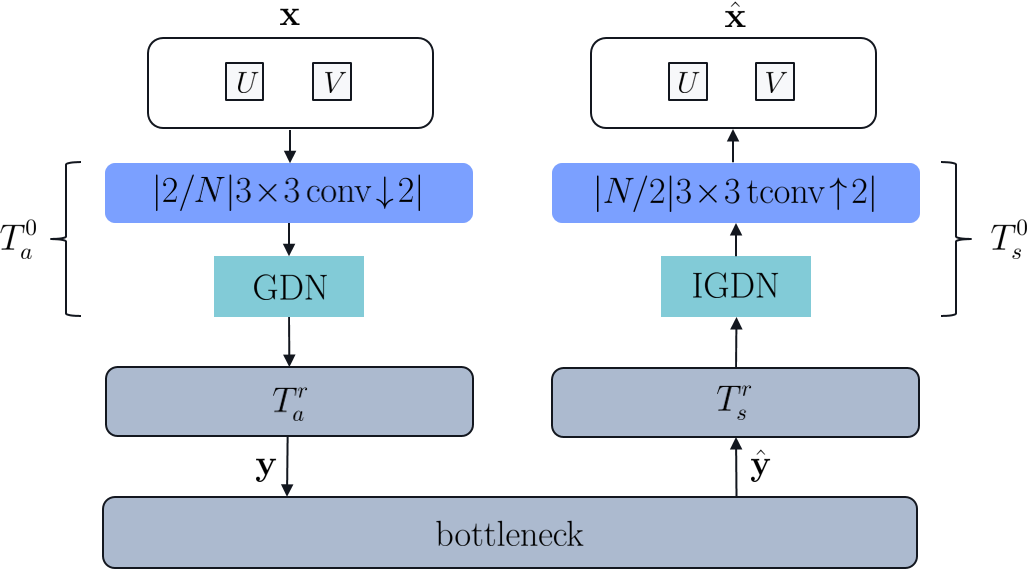}} 
\caption{Network architectures for separate-channel coding of (a) luma and (b) chroma components: Both networks are based on Fig.~\ref{fig:google_transfrom_network}, where the same bottleneck and transform layers $T_a^r$ and $T_s^r$ are used. For (a) and (b), the common difference (compared to Fig.~\ref{fig:google_transfrom_network}) is the number of channels at the encoder-input and decoder-output in $T_a^0$ and $T_s^0$. The network in (b) further uses a smaller kernel in $T_a^0$ and $T_s^0$.}    
\label{fig:method_separate}
\end{figure}

\begin{figure}[!t]
\centering
    {\includegraphics[width=0.8\columnwidth]{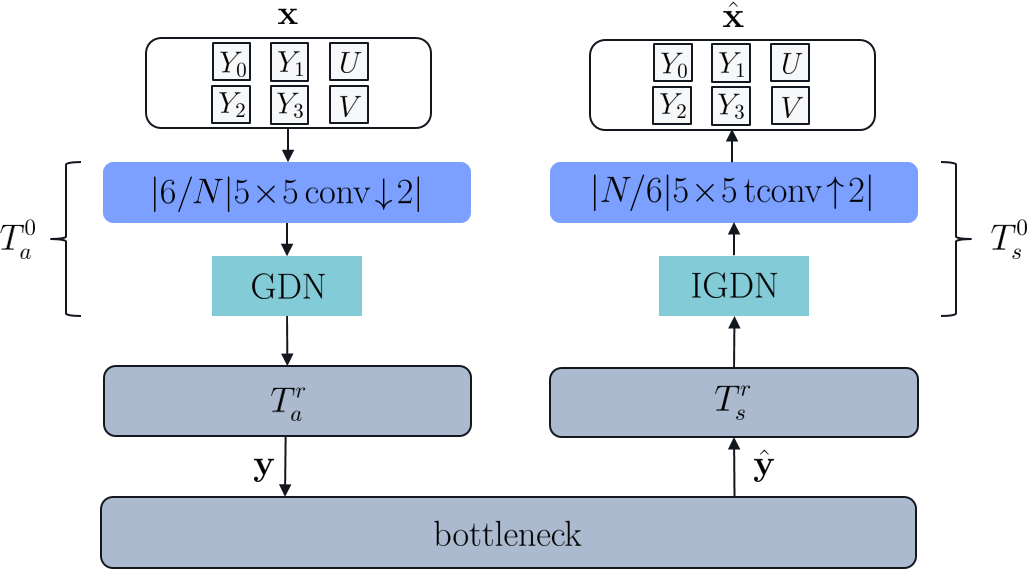}} 
\caption{Network architectures for six-channel coding: Both networks are based on Fig.~\ref{fig:google_transfrom_network}, where the same bottleneck and transform layers $T_a^r$ and $T_s^r$ are used, and the only difference (compared to Fig.~\ref{fig:google_transfrom_network}) is the number of channels at the encoder-input and decoder-output in $T_a^0$ and $T_s^0$.}    
\label{fig:method_six}
\end{figure}

Based on the input-output channel alignment approach, 
we introduce the following two coding schemes.
\subsubsection{Separate-channel Coding} 
In separate coding scheme, Y, U and V components are split channel-wise as shown in Fig.~\ref{fig:method_separate} so that luma (Y) and chroma (U and V) channels are coded separately. This coding approach requires training two separate networks, where the only difference in each network with respect to Fig.~\ref{fig:google_transfrom_network} is the number of encoder-input and decoder-output channels. 
Accordingly, for coding luma component, the network (Fig.~\ref{fig:separate_ch_y}) uses a single encoder-input/decoder-output channel, and the other network (Fig.~\ref{fig:separate_ch_uv}) has two encoder-input/decoder-output channels is used for coding chroma (U and V components). 

In principle, coding chroma components is easier (i.e., requires much less bitrate) than coding luma, so a smaller $3 \times 3$ convolution kernel (Fig.~\ref{fig:separate_ch_uv}) is used with little or no loss in chroma coding efficiency as compared to using a $5 \times 5$ kernel. Similarly, in block-based video coding standards, more sophisticated compression tools are utilized for coding luma than for coding chroma components.

\subsubsection{Six-channel Coding}
For six-channel coding, the luma component (Y) is spatially split into four channels by matching the dimensions of chroma components. The resulting four luma channels are then stacked with two chroma channels, so the six-channel format is constructed as shown in Fig.~\ref{fig:method_six}. Unlike separate-channel coding, this approach only requires training a single network, yet the number of encoder-input/decoder-output channels to the network is increased to six. 

\begin{figure}[!t]
\centering
    {\includegraphics[width=0.6\columnwidth]{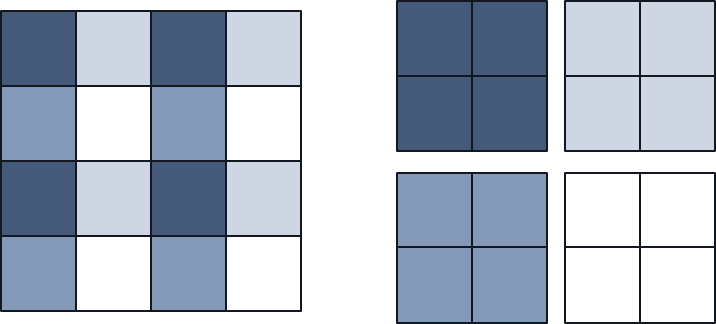}}
\caption{An example of splitting used to construct four channels by sampling a $4 \times 4$ block, where out of a channel with 16 pixels four channels with each consisting of 4 samples are created.}
\label{fig:six_ch_split}
\end{figure}

In this paper, we specifically apply the channel splitting method depicted in Fig.~\ref{fig:six_ch_split} on luma component to construct four channels where each channel is constructed by sampling every other luma pixel in each direction.

\begin{figure}[!t]
\centering
    {\includegraphics[width=\columnwidth]{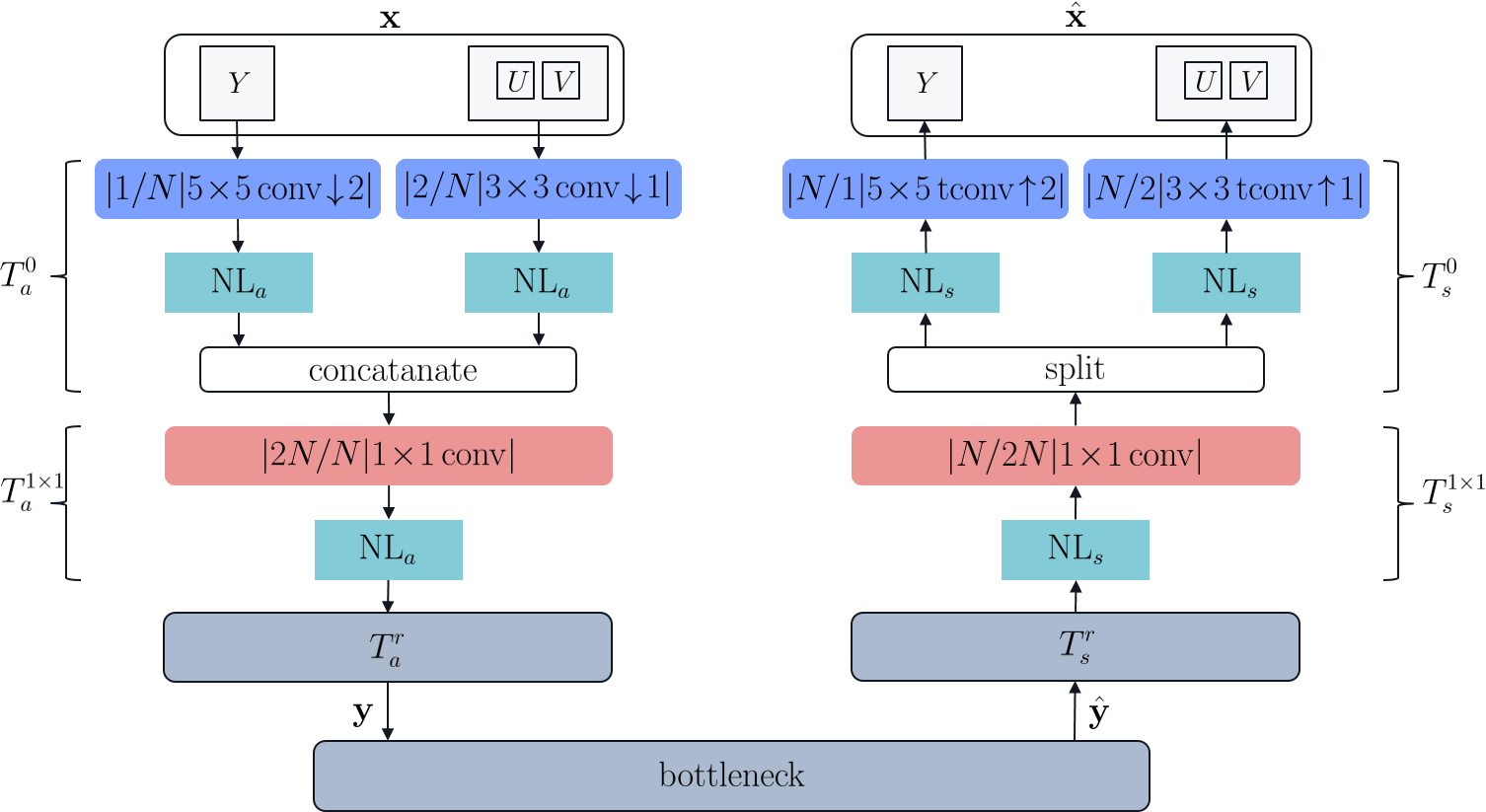}} 
\caption{The proposed architecture built by integrating (i) the branched network structures $T_a^0$ and $T_s^0$, (ii) additional $1 \times 1$ convolutional layers denoted by $T_a^{1\times1}$ and $T_s^{1\times1}$ and (iii) different choices of nonlinear operators $\mathrm{NL}_a$ and $\mathrm{NL}_s$. The remaining transform network layers denoted by $T_a^r$ and $T_s^r$ follow the same structure in Fig.~\ref{fig:google_transfrom_network}.}    
\label{fig:overall_architecture}
\end{figure}

\subsection{Proposed Transform Network Architecture} 
\subsubsection{Building blocks of the proposed architecture}
The proposed architecture is built upon the 
transform network in Fig.~\ref{fig:google_transfrom_network}, 
where we incorporate the three building blocks to construct our architecture shown in Fig.~\ref{fig:overall_architecture}, discussed in the following.

\noindent{\bf{Branched network structure:}} 
The {branched network} denoted as $T_a^0$ in Fig.~\ref{fig:overall_architecture} 
is used as the first step for transforming luma and chroma channels separately at the encoder side. The luma branch uses a 5$\times$5 
convolutional layer with downsampling by 2, that is identical to the first layer in Fig.~\ref{fig:google_transfrom_network},   
whereas the chroma branch does not apply any downsampling (i.e., stride is set to 1) to equalize the output channel dimensions of both branches, which are concatenated after a nonlinear activation. At the decoder counterpart, the corresponding branched network with transposed convolutional layer $T_s^0$ (see in Fig.~\ref{fig:overall_architecture}) is applied as the last step of reconstruction. Note that, as compared to the kernel in luma branch, chroma branch uses a smaller (3$\times$3) kernel as in Fig.~\ref{fig:separate_ch_uv} since U and Y components are often easier to compress and it is generally sufficient to use a kernel with smaller support. Hence, the branched network structure naturally allows us to use different types of convolutions (with sampling or no sampling) needed to align the channel dimensions of luma and subsampled chroma for their joint coding after channel concatenation. In addition, it allows to capture different signal characteristics in luma and chroma channels by using different kernel sizes at the initial stage of the analysis transform network.

\noindent {\bf{Cross-channel transformation:}} 
{1$\times$1 convolutional layer} is used to combine the output of branched network carrying luma and chroma information separately at the encoder side. From the compression perspective, a 1$\times$1 convolution can be viewed as a cross-channel transformation/prediction process that can exploit correlation between luma and chroma channels. As shown in Fig.~\ref{fig:overall_architecture}, 1$\times$1 convolution ($T_a^{1 \times 1}$) at the encoder side halves the number of channels that are concatenated in the previous layer. At the decoder part, the corresponding 1$\times$1 convolution ($T_s^{1 \times 1}$) 
doubles the number channels before splitting them into two branches. 
In the literature, 1$\times$1 convolution is originally proposed for image classification \cite{Lin:2014:1x1conv}, and it is also used in part of the \emph{Inception} architecture \cite{szegedy:2014:inception} achieving high classification performances. 

\noindent {\bf{Activation functions:}} The parametric rectified linear unit (PReLU) \cite{He:2015:PReLU} is one of the activation functions used in our design to adjust the scaling of negative filter responses with a learnable (slope) parameter $a$. For compression purposes, PReLU is better suited as compared to ReLU and leaky-ReLU for compression, because (i) ReLU sets any negative filter responses to zero, that is undesirable to lose  information at transform layers before the quantization and entropy coding steps, and (ii) having different nonlinear functions (with a trainable slope $a$) between layers provides a better adaptation as compared to leaky-ReLU with a fixed slope. 
Besides, pointwise nonlinear operators such as PReLU is significantly less complex than GDN, which apply an additional  cross-channel transformation on filter responses before the nonlinear normalization.  

\subsubsection{Variants of the proposed architecture}
\begin{figure}[!t]
\centering
    \subfloat[GDNs are used for all the activations.\label{fig:arc1_gdns}]{\includegraphics[width=0.98\columnwidth]{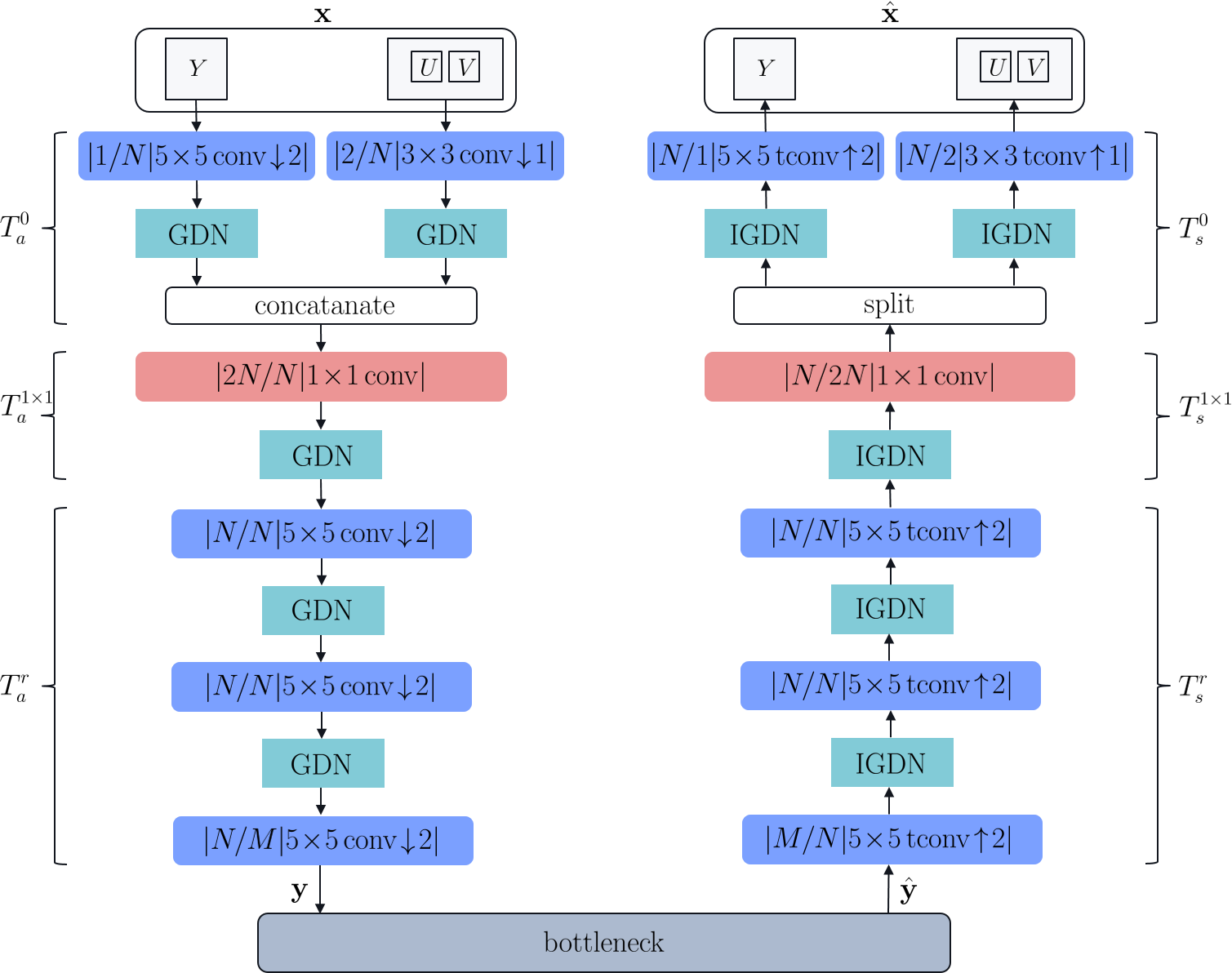}} \\
    \subfloat[Two PReLUs are used adjacent to 1$\times$1 convolutions.\label{fig:arc2_prelu}]{\includegraphics[width=0.98\columnwidth]{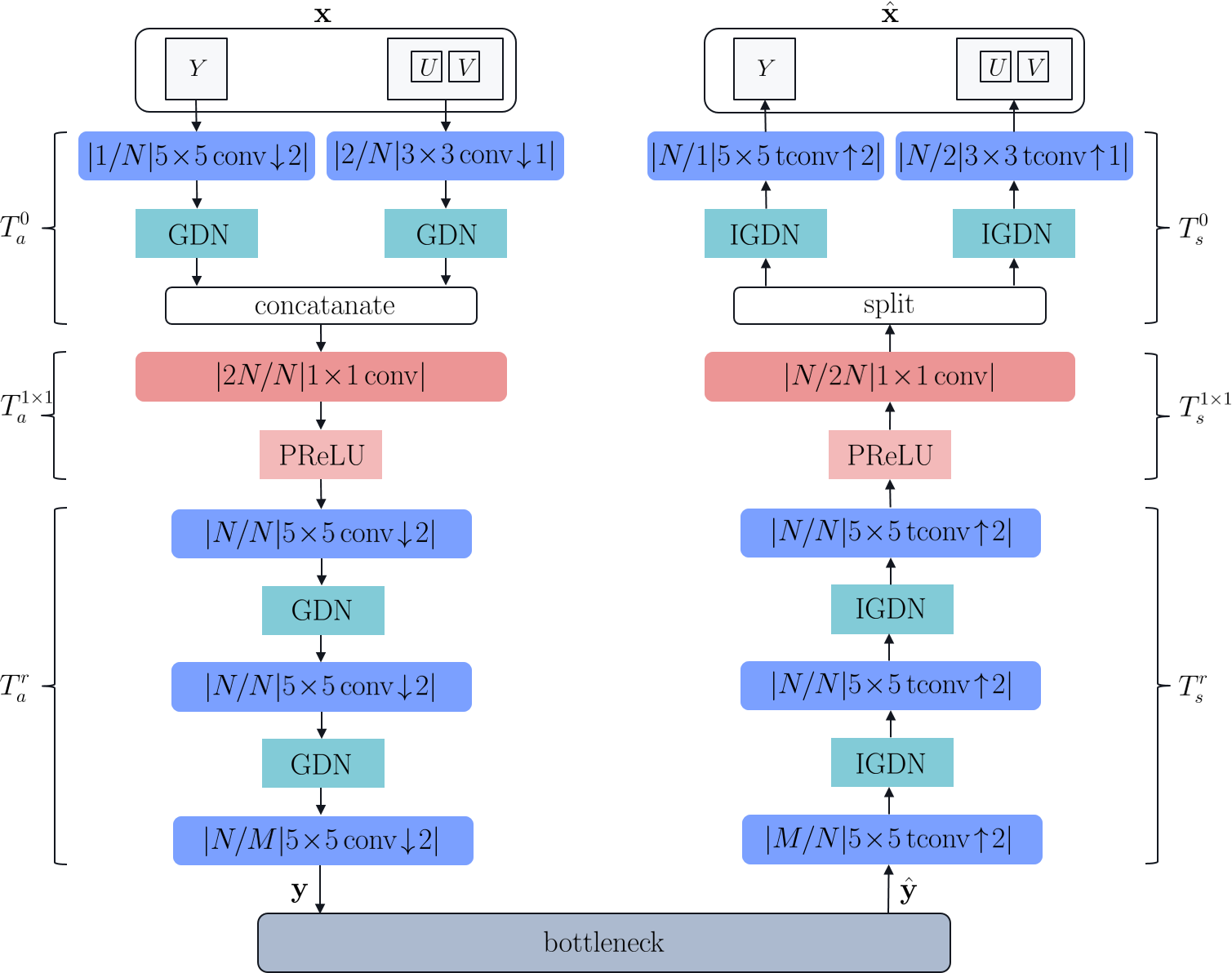}} \\
    \subfloat[PReLUs are used for all the activations (i.e., all GDNs are removed). \label{fig:arc3_allprelu}]{\includegraphics[width=0.98\columnwidth]{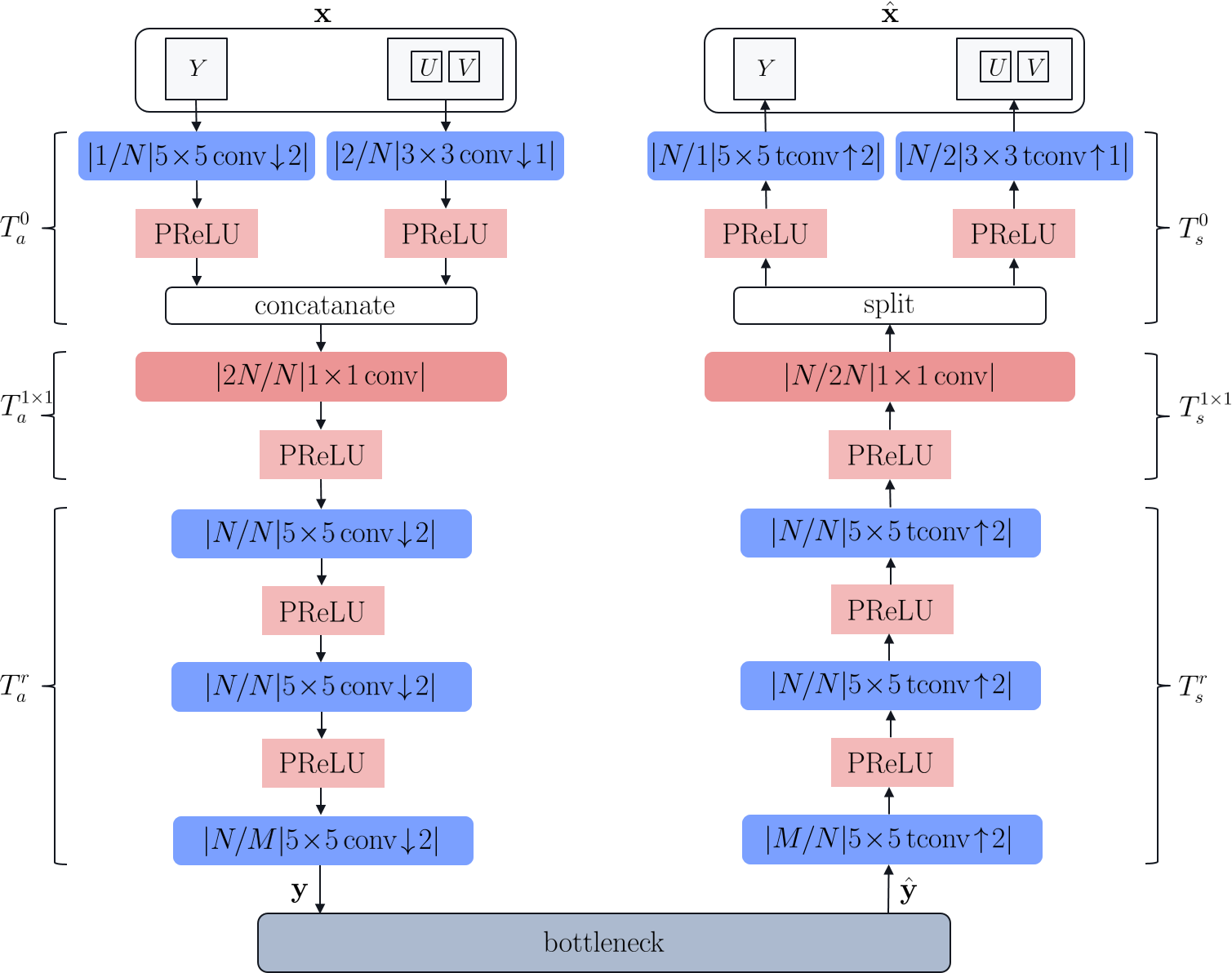}}
\caption{Three variants of the proposed architecture using (a) GDNs only, (b) a combination of GDNs and PReLUs and (c) PReLUs only.}
\label{fig:three_variants}
\end{figure}

As a nonlinear operator, GDN has been extensively used in recent studies on {DLEC} \cite{Balle:2018:ICLR:ScaleHyper,MinnenBT:2018:NeurIPS:MSHyper,Minnen:2020:ICIP:Charm} 
because of its effective cross-channel normalization whose coding performance is shown to be better than ReLU and leaky-ReLU on RGB data \cite{Balle:2018:GDN}.
Noting that 1$\times$1 convolutions in the proposed architecture also serve as cross-channel transformations, we introduce variants of the proposed transform network architecture (with different choices of activation functions) to investigate the effect of GDNs and 1$\times$1 convolutions on coding performance. 
Specifically, in this paper, three transform network variants shown in Fig.~\ref{fig:three_variants} are evaluated. The first transform network in Fig.~\ref{fig:arc1_gdns} uses GDNs for all activations, while the other two variants gradually replace GDNs with PReLUs. Respectively, the network in Fig.~\ref{fig:arc2_prelu} replaces two GDNs adjacent to 1$\times$1 convolutional layers with PReLUs at both encoder and decoder, and Fig.~\ref{fig:arc3_allprelu} only uses PReLUs as activation functions.

Table \ref{table:complexity} shows the complexity (or size) of each transform network solution in terms of the number of trainable parameters. Note that the best performing network in Fig.~\ref{fig:arc3_allprelu} has the fewest number of parameters since all GDNs are replaced with simpler PReLUs. As compared to Fig.~\ref{fig:arc3_allprelu}, the networks in Figs.\ref{fig:arc2_prelu} and \ref{fig:arc1_gdns} have 296,472 and 370,590 more trainable parameters, respectively. 
Since the separate-channel coding employs two separate networks for luma and chroma channels, it nearly doubles the number of parameters (i.e., about 14 million) over all other solutions introduced in this paper. 

\begin{table}[!t]
\centering
\caption{Number of trainable parameters in transform networks}
\renewcommand{\arraystretch}{1.1}
\label{table:complexity}
\begin{tabular}{|c|c|c|c|c|} \hline
Separate & Six-channel &  Fig.~\ref{fig:arc1_gdns} &  Fig.~\ref{fig:arc2_prelu} & Fig.~\ref{fig:arc3_allprelu}  \\ \hline \hline
14,004,411 & 7,014,690 & 7,306,927 & 7,232,809 & 6,936,337  \\ \hline
\end{tabular}
\end{table}

\section{Results}
\label{sec:results}
\subsection{Simulation Setup, Training and Testing}
\label{subsec:sim_setup}
The transform network solutions discussed in Section \ref{sec:dnn_archs} are implemented, trained and tested on CompressAI framework \cite{begaint:2020:compressai}, which is a deep learning library based on PyTorch for end-to-end image/video compression. Throughout the experiments, the number of channels in the transform network and bottleneck are set as $N=192$ and $M=320$, respectively. 
For entropy coding at the bottleneck, the arithmetic coding engine in \cite{duda:2014:ANS} called asymmetric numeral systems (ANS) is employed. As for the entropy model, the mean-scale hyperprior network in \cite{MinnenBT:2018:NeurIPS:MSHyper} is integrated. {The tested DLEC codec operates in double precision, and the ANS entropy coder has an internal bitdepth of 16 bits.}

For training dataset, we generated YUV 4:2:0 video content by converting the 
448$\times$256 RGB videos in Vimeo-90k Triplet dataset \cite{dataset:vimeo-90k-triplet} into YUV 4:2:0 format using \emph{ffmpeg} software \cite{software:ffmpeg}. 
The resulting YUV 4:2:0 frames (with dimensions 448$\times$256 for luma and 224$\times$128 for chroma components) are used for training, where the batch size is set to 8. Each model is trained for 4 million steps using Adam algorithm \cite{Kingma:2015:iclr:adam}.
Table \ref{table:training} presents some details about the training. 

\begin{table}[!t]
\centering
\caption{Detailed information on training} \label{table:training}
\renewcommand{\arraystretch}{1.1}
\label{table:result}
\begin{tabular}{|c|c|} \hline
Dataset    &   Vimeo-90k Triplet \cite{dataset:vimeo-90k-triplet} (converted to YUV 4:2:0)  \\ \hline
Input size &   448$\times$256 for Y, 224$\times$128 for U and V   \\ \hline
Batch size  &   8 frames (in YUV 4:2:0 format)  \\ \hline
Optimizer & Adam algorithm \cite{Kingma:2015:iclr:adam} 
\\ \hline
Number steps  &   4 million steps \\ \hline
Learning rate ($r$) & initially $r = 10^{-4}$ then $r = 10^{-5}$ after 2M steps \\ \hline

\end{tabular}
\end{table}

During training, the following loss function (i.e., RD cost based on (\ref{eqn:rd_cost_neural})) with component-wise weighted mean-squared error (MSE) in the distortion term:
\begin{equation} \label{eqn:loss_initial}
    L = R + \beta \underbrace{(8\,\text{MSE}_Y + 2\,\text{MSE}_U + 2\,\text{MSE}_V)/12}_{D}   \\
\end{equation}
where $D$ is a weighted combination of MSEs obtained from Y, U and V components (denoted by $\text{MSE}_Y$, $\text{MSE}_U$ and $\text{MSE}_V$, respectively). Note that the weighted terms are proportional to the size of each component (in YUV 4:2:0 format), so that contribution of Y component is 4 times that of U or V. 
For each DLEC approach, five separate models are trained for different $\beta$ parameters ${0.005, 0.01, 0.025, 0.1, 0.2}$ to evaluate the coding performance at different rate points.

For testing, the video sequences in \emph{common test conditions (CTC)} \cite{Bossen:18:ctc} defined for VVC standardization activities are used as our primary test dataset, which consists of diverse classes of raw (uncompressed) video sequences in  {8 or 10-bit} YUV 4:2:0 format. The coding performance is evaluated by averaging bitrate reduction over five classes of video sequences, where classes A1 and A2 consist of 4K UHD (3840$\times$2160) sequences, class B has HD (1920$\times$1080) content, and sequences with 832$\times$480 and 1280$\times$720 resolutions are included in classes C and E, respectively.
Since this paper focuses on intra-frame coding, our experimental evaluation is performed according to the \emph{all-intra} configurations defined in CTC, where every eighth frame of each sequence is coded throughout the evaluations. Moreover, the trained models are tested on widely used Kodak image dataset \cite{dataset:Kodak} after they are converted to {8-bit} YUV 4:2:0 format by using \emph{ffmpeg} \cite{software:ffmpeg}. 

The compression performance is measured by Bjontegaard-delta rate (BD-rate or BDR) \cite{bd_metric}, which is the metric used in HEVC and VVC standardization \cite{Bossen:18:ctc,Bossen:13:hevc_ctc}. 
Essentially, BDR quantifies the total bitrate reduction achieved by the tested codec over a reference codec, called anchor, while maintaining the same quality level measured by an objective metric. As in CTC, peak signal-to-noise ratio (PSNR) is calculated to measure quality component-wise (separately for Y, U and V components), and the BDR is calculated based on different bitrate operating points, which are obtained by coding at different QPs in traditional codecs. 
In evaluation of DLEC approaches, the selected five $\beta$ values 0.005, 0.01, 0.025, 0.1 and 0.2 roughly correspond to bitrates achieved with HEVC at QPs 42, 37, 32, 27 and 22, respectively, according to our experiments.

As in CTC, the compression performances are evaluated in terms of component-wise BD-rates measured separately for Y, U and V (denoted by Y-BDR, U-BDR and V-BDR), while Y-BDR (luma) being by far the most important component as compared to U-BDR and V-BDR (chroma). 
Based on our extensive experiments, we observe that, on average, a 1\% Y-BDR improvement is approximately equivalent to 12\% combined U-BDR and V-BDR gain (i.e., about 6\% gain for each U and V). 
Accordingly, we introduce the following combined BD-rate (CBDR) as another metric quantifying overall coding performance:
\begin{equation}
\label{eqn:CBDR}
\text{CBDR} = (12 \, \text{Y-BDR} + \text{U-BDR} + \text{V-BDR})/{14}
\end{equation}
where the weights for each component (i.e., 12/14, 1/14 and 1/14 for Y, U and V, respectively) are chosen based on our empirical study. 
In our results, CBDR is also reported together with component-wise BD-rates.

\subsection{Comparison of various DLEC solutions for YUV 4:2:0}
\label{subsec:e2e}
Tables \ref{table:result_Sep_vs_6ch}--\ref{table:result_Sep_vs_joint_pReLU_allRemoved} present the coding performance of DLEC solutions discussed in Section \ref{sec:dnn_archs} on CTC sequences and Kodak image dataset. 
In each table, the tested coding scheme is evaluated by benchmarking against the separate-channel coding as anchor, and component-wise BD-rates and CBDR are reported, where a negative BDR value means that tested scheme provides better compression than the anchor. 

Table \ref{table:result_Sep_vs_6ch} shows that between two input-output alignment solutions, the six-channel coding approach outperforms separate-channel coding on both CTC sequences and Kodak images. Based on the average CBDR results, the six-channel coding achieves 0.16\% gain on CTC sequences and 1.23\% gain on Kodak dataset over separate coding, while only on class E sequences the six-channel coding underperforms by 3.73\%.

As compared to Table \ref{table:result_Sep_vs_6ch}, the results in Tables \ref{table:result_Sep_vs_joint_GDN}, \ref{table:result_Sep_vs_joint_pReLU} and \ref{table:result_Sep_vs_joint_pReLU_allRemoved} 
clearly demonstrate the benefit of the proposed transform network architecture over the input-output channel alignment solutions, where all variants of the proposed architecture (in Fig.~\ref{fig:three_variants}) significantly outperform the separate-channel coding with 
more than 3\% average CBDR reduction on CTC sequences 
and more than 4.5\% CBDR reduction on Kodak dataset.

\begin{table}[!t]
\centering
\caption{
Test: Six-channel coding, Anchor: Separate coding}
\renewcommand{\arraystretch}{1.1}
\label{table:result_Sep_vs_6ch}
\begin{tabular}{|c|c|c|c|c|} \hline
{Test Dataset}             &    Y-BDR       &     U-BDR    &    V-BDR      &      CBDR  \\ \hline \hline
{CTC-Class A1}           &    0.28\%      & 	-6.70\%    &  -15.58\%     &    -1.35\% \\ \hline
{CTC-Class A2}           &   -2.65\%      & 	-2.70\%    &    1.39\%     &    -2.36\% \\ \hline
{CTC-Class B}            &   -0.28\%      & 	 3.26\%    &   -6.22\%     &    -0.45\% \\ \hline
{CTC-Class C}            &    0.62\%      & 	-2.49\%    &   -7.05\%     &    -0.15\% \\ \hline
{CTC-Class E}            &    2.23\%      & 	16.49\%    &    8.92\%     &     3.73\% \\ \hline \hline
\bf{CTC (Overall)}  	   &    0.04\%      & 	 1.53\%    &   -4.18\%     &    -0.16\% \\ \hline \hline
\bf{Kodak (Overall)} 	   &    0.01\%      &  -11.51\%    &   -5.86\%	   &    -1.23\% \\ 
\hline
\end{tabular}
\end{table}

\begin{table}[!t]
\centering
\caption{Test: Transform network in Fig.~\ref{fig:arc1_gdns}, Anchor: Separate coding}
\renewcommand{\arraystretch}{1.1}
\label{table:result_Sep_vs_joint_GDN}
\begin{tabular}{|c|c|c|c|c|} \hline
{Test Dataset}             &    Y-BDR     &    U-BDR     &   V-BDR      &   CBDR   \\ \hline \hline
{CTC-Class A1}           &   -7.94\%    &   -21.22\%   &  -23.11\%    &  -9.97\% \\ \hline
{CTC-Class A2}           &   -6.89\%    &    -4.82\%   &    2.42\%    &  -6.07\% \\ \hline
{CTC-Class B}            &    1.26\%    &     3.35\%   &   -4.69\%    &   0.99\% \\ \hline
{CTC-Class C}            &   -3.60\%    &    -5.11\%   &   -6.77\%    &  -3.94\% \\ \hline
{CTC-Class E}            &   -0.86\%    &    16.03\%   &    8.42\%    &   1.01\% \\ \hline \hline
\bf{CTC (Overall)}         &   -3.07\%    &    -1.87\%   &   -4.85\%    &  -3.11\% \\ \hline \hline
\bf{Kodak (Overall)}       &   -2.95\%    &   -18.49\%   &  -12.11\%	&  -4.71\% \\ 
\hline
\end{tabular}
\end{table}

\begin{table}[!t]
\centering
\caption{Test: Transform network in Fig.~\ref{fig:arc2_prelu}, Anchor: Separate coding}
\renewcommand{\arraystretch}{1.1}
\label{table:result_Sep_vs_joint_pReLU}
\begin{tabular}{|c|c|c|c|c|} \hline
{Test Dataset}             &    Y-BDR     &      U-BDR      &    V-BDR     &    CBDR  \\ \hline \hline
{CTC-Class A1}           &   -9.36\%    &     -25.09\%    &  -23.67\%    & -11.50\% \\ \hline
{CTC-Class A2}           &   -7.08\%    & 	 -5.48\%    &    1.35\%    &  -6.36\% \\ \hline
{CTC-Class B}            &   -6.32\%    & 	  0.80\%    &   -9.22\%    &  -6.01\% \\ \hline
{CTC-Class C}            &   -5.24\%    & 	 -7.20\%    &   -7.06\%    &  -5.51\% \\ \hline
{CTC-Class E}            &   -5.49\%    & 	 11.79\%    &    1.70\%    &  -3.75\% \\ \hline \hline
\bf{CTC (Overall)}         &   -6.57\%    & 	 -4.51\%    &   -7.57\%    &  -6.50\% \\ \hline \hline
\bf{Kodak (Overall)}       &   -4.41\%    &     -16.40\%    &   -9.86\%	   &  -5.66\% \\ \hline
\end{tabular}
\end{table}

\begin{table}[!t]
\centering
\caption{Test: Transform network in Fig.~\ref{fig:arc3_allprelu}, Anchor: Separate coding}
\renewcommand{\arraystretch}{1.1}
\label{table:result_Sep_vs_joint_pReLU_allRemoved}
\begin{tabular}{|c|c|c|c|c|} \hline
{Test Dataset}            &    Y-BDR    &     U-BDR    &    V-BDR    &     CBDR    \\ \hline \hline
{CTC-Class A1}           &   -8.46\%    &   -29.34\%    &  -23.48\%    &  -11.02\%   \\ \hline
{CTC-Class A2}           &   -6.81\%    &    -7.73\%    &   -1.89\%    &   -6.53\%   \\ \hline
{CTC-Class B}            &   -6.85\%    &     5.19\%    &  -11.18\%    &   -6.30\%   \\ \hline
{CTC-Class C}            &   -5.50\%    &    -9.68\%    &  -12.08\%    &   -6.27\%   \\ \hline
{CTC-Class E}            &   -6.81\%    &     5.08\%    &    5.70\%    &   -5.07\%   \\ \hline \hline
\bf{CTC (Overall)}   &   -6.81\%    &    -6.04\%    &   -9.07\%    &   -6.91\%   \\ \hline \hline
\bf{Kodak (Overall)} &   -4.87\%    &   -16.25\%    &  -11.84\%	   &   -6.18\%   \\ \hline
\end{tabular}
\end{table}

Among the variants of our proposed architecture, 
the best performing solution (based on the average BD-rate results in Tables \ref{table:result_Sep_vs_joint_GDN}, \ref{table:result_Sep_vs_joint_pReLU} and \ref{table:result_Sep_vs_joint_pReLU_allRemoved}) 
is the transform network shown in Fig.~\ref{fig:arc3_allprelu}, 
which only uses PReLU activations and no GDNs.  
Specifically, the corresponding results in Table \ref{table:result_Sep_vs_joint_pReLU_allRemoved} show that 6.91\% CBDR reduction is achieved over separate coding on CTC sequences, while the variants depicted in Figs.\ref{fig:arc2_prelu} and \ref{fig:arc1_gdns} result in lower coding gains with 6.50\% (see Table \ref{table:result_Sep_vs_joint_pReLU}) 
and 3.11\% (see Table \ref{table:result_Sep_vs_joint_GDN}) CBDR savings, respectively. Note that the transform network in Fig.~\ref{fig:arc3_allprelu} is the best performing solution not only in CBDR but also in terms of Y-BDR, U-BDR and V-BDR metrics on both CTC sequences and Kodak images.   

The compression results demonstrate that layers of cross-channel normalizations 
via GDNs are not necessary, and they can be replaced with simpler PReLUs as the proposed 1$\times$1 convolutions in our architecture cover the benefit of 
GDNs and provides further coding improvements with their effective 
cross-component transformation capabilities\footnote{1$\times$1 convolutions can also be interpreted as cross-channel predictions that  can reduce cross-channel redundancies in the data to provide coding gains.}.

\subsection{Comparison of the proposed solution against traditional block-based codecs}
\label{subsec:compare_codecs}
Table \ref{table:result_HEVC_vs_joint_pReLU_allRemoved} compares the 
compression performance of proposed transform network in Fig.~\ref{fig:arc3_allprelu} against the HEVC standard in terms of BD-rates\footnote{Based on the BD-rate results in Section \ref{subsec:e2e}, the transform network in Fig.~\ref{fig:arc3_allprelu} is the best performing solution among all introduced in this paper.}. 
The HEVC coding results are obtained by compressing both CTC sequences \cite{Bossen:18:ctc} and Kodak images \cite{dataset:Kodak} using the HEVC reference software (HM-16.20)\footnote{HM is the reference software used as a baseline in standardization activities and many research articles, as it reflects the state-of-the-art compression capabilities of HEVC. However, product-driven compression tools such as \emph{ffmpeg} and \emph{x265} can lead to a significantly lower coding efficiency than HM. For example, \cite{HHI:21:vvenc_vvdec} shows that encoding with \emph{x265} under the complex \emph{placebo} presetting is about 15\% less efficient than HM in terms of BD-rate.} under \emph{all-intra} coding configurations \cite{Bossen:13:hevc_ctc} at five QPs, 42, 37, 32, 27 and 22. 
The table shows that our proposed DLEC solution outperforms HEVC on average by 7.35\% in CBDR and 12.61\% in Y-BDR, yet it leads to a considerable coding loss in chroma components especially in class A1 and A2 sequences. On the other hand, the coding performance on Kodak dataset is relatively more balanced across components, where the overall coding gain of proposed solution is 9.28\% over HEVC.

In order to balance the luma and chroma coding performance on CTC sequences, we retrained the  parameters of the proposed architecture in Fig.~\ref{fig:arc3_allprelu} by changing the weights of the distortion term in the loss function in (\ref{eqn:loss_initial}) as\footnote{The training is performed by following the same procedure described in Section \ref{subsec:sim_setup}. Only the distortion term in (\ref{eqn:loss_initial}) is replaced with (\ref{eqn:loss_reweighted}).}: 
\begin{equation} \label{eqn:loss_reweighted}
    L_{\mathsf{new}} = R + \beta \underbrace{(6\,\text{MSE}_Y + 3\,\text{MSE}_U + 3\,\text{MSE}_V)/12}_{D}   \\
\end{equation}
which increases weights for U and V components 
so that the loss function penalizes the distortion on chroma components more than it is penalized in (\ref{eqn:loss_initial}). The coding performance of the retrained models over HEVC are shown in Table \ref{table:result_HEVC_vs_joint_pReLU_allRemoved_reweighted}, where the new models significantly improved chroma coding gains (U-BDR and V-BDR) with some loss in luma component (Y-BDR) as expected. In terms of the overall CBDR results (see in Table \ref{table:result_HEVC_vs_joint_pReLU_allRemoved_reweighted}), the retrained models provide a better compression trade-off where the overall CBDR reduction is increased to 10.21\% on CTC sequences and to 11.63\% on Kodak images. It is important to note that the proposed new architecture supporting YUV 4:2:0 format allows us to balance (tune) the coding gains between Y, U and V components considering the human visual system which is more sensitive to distortion in Y components, while such balancing is not feasible for networks working with RGB data.

\begin{figure}[!t]
\centering
\subfloat[Results on CTC video sequences \label{fig:rd_curve_CTC}]
{\includegraphics[width=0.95\columnwidth,trim={0 0 0 1.4cm},clip]{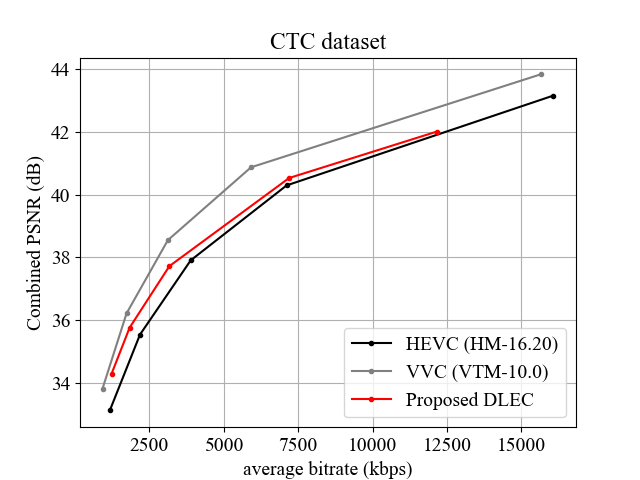}} \\
\subfloat[Results on Kodak image dataset \label{fig:rd_curve_kodak}]
{\includegraphics[width=0.95\columnwidth,trim={0 0 0 1.4cm},clip]{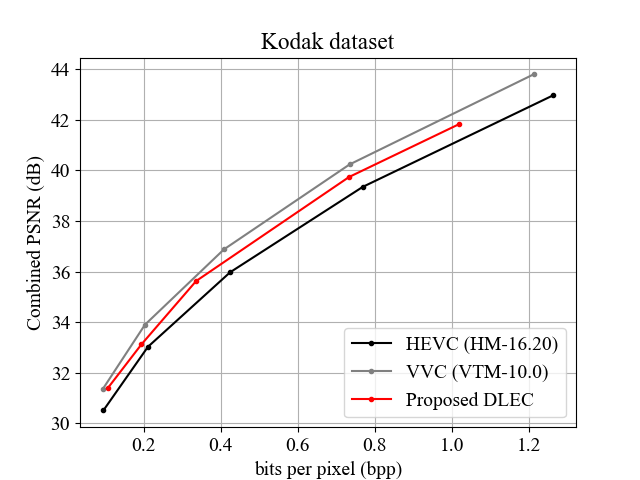}}
\caption{Rate-distortion (RD) curves corresponding to the CBDR results in Table \ref{table:result_vvc_proposed_hevc} tested on (a) CTC video sequences and (b) Kodak images. The average rates are reported in kilobits per second (kbps) for the CTC video dataset and in bits per pixel (i.e., total bits divided by the resolution) for Kodak image dataset.}
\label{fig:rd_curves}
\end{figure}

 Table \ref{table:result_vvc_proposed_hevc} {summarizes} 
 the {overall} CBDR reduction achieved by VVC\footnote{VTM reference software (version 10.0) is used for coding at five QPs, 42, 37, 32, 27 and 22. {Table \ref{table:result_HEVC_vs_VTM} shows per-class and per-component coding performance of VVC over HEVC.}} over HEVC together with the proposed architecture with retrained (i.e., luma-chroma balanced) models, where the proposed DLEC solution is about 16\% less efficient than VVC in coding CTC sequences, and it is about 9\% less efficient in coding Kodak images. The corresponding RD-curves for HEVC, VVC and proposed DLEC solution with retrained models are presented in Fig.~\ref{fig:rd_curves}, which shows that the proposed DLEC is about half-way between HEVC and VVC in terms of the average coding performance (CBDR). The RD-curves also show that the proposed DLEC performs relatively better at lower bitrates especially in coding CTC sequences (e.g., bitrates less than 5000 kbps in Fig.~\ref{fig:rd_curve_CTC}).
 
\begin{table}[!t]
\centering
\caption{Test: Proposed solution (Fig.~\ref{fig:arc3_allprelu}), Anchor: HEVC (HM-16.20)}
\renewcommand{\arraystretch}{1.1}
\label{table:result_HEVC_vs_joint_pReLU_allRemoved}
\begin{tabular}{|c|c|c|c|c|} \hline
{Test Dataset}             &    Y-BDR      &     U-BDR     &    V-BDR      &     CBDR    \\ \hline \hline
{CTC-Class A1}           &   -18.12\%    &    74.46\%    &  -10.32\%     &  -10.95\%   \\ \hline
{CTC-Class A2}           &   -19.26\%    &    70.31\%    &   79.23\%     &   -5.83\%   \\ \hline
{CTC-Class B}            &   -10.16\%    &    23.22\%    &   -4.30\%     &   -7.36\%   \\ \hline
{CTC-Class C}            &    -6.20\%    &    21.44\%    &   10.98\%     &   -3.00\%   \\ \hline
{CTC-Class E}            &   -13.08\%    &    12.85\%    &  -10.92\%     &  -11.07\%   \\ \hline \hline
\bf{CTC (Overall)}         &   -12.61\%    &    37.48\%    &   10.91\%     &   -7.35\%   \\ \hline \hline
\bf{Kodak (Overall)}       &    -7.62\%    &   -24.35\%    &  -14.11\%	   &   -9.28\%   \\ \hline
\end{tabular}
\end{table}

\begin{table}[!t]
\centering
\caption{Test: Proposed solution (Fig.~\ref{fig:arc3_allprelu}) trained with $L_{\mathsf{new}}$ in (\ref{eqn:loss_reweighted}), Anchor: HEVC (HM-16.20)}
\renewcommand{\arraystretch}{1.1}
\label{table:result_HEVC_vs_joint_pReLU_allRemoved_reweighted}
\begin{tabular}{|c|c|c|c|c|} \hline
{Test Dataset}             &     Y-BDR     &     U-BDR     &    V-BDR     &     CBDR    \\ \hline \hline
{CTC-Class A1}           &   -14.39\%    &    22.43\%    &  -35.80\%    &  -13.29\%   \\ \hline
{CTC-Class A2}           &   -16.56\%    &    22.44\%    &   26.70\%    &  -10.68\%   \\ \hline
{CTC-Class B}            &    -6.22\%    &   -15.64\%    &  -32.96\%    &   -8.81\%   \\ \hline
{CTC-Class C}            &    -4.62\%    &   -12.42\%    &  -20.42\%    &   -6.31\%   \\ \hline
{CTC-Class E}            &   -12.04\%    &   -15.19\%    &  -39.53\%    &  -14.23\%   \\ \hline \hline 
\bf{CTC (Overall)}         &    -9.92\%    &    -2.16\%    &  -21.80\%    &  -10.21\%   \\ \hline \hline
\bf{Kodak (Overall)}       &    -5.95\%    &   -48.21\%    &  -43.20\%	  &  -11.63\%   \\ \hline
\end{tabular}
\end{table}

\begin{table}[!t]
\centering
\caption{Coding performance of VVC (VTM-10.0) and proposed solution (Fig.~\ref{fig:arc3_allprelu}) trained with $L_{\mathsf{new}}$ in (\ref{eqn:loss_reweighted}) over HEVC (HM-16.20) anchor in terms of CBDR.}
\renewcommand{\arraystretch}{1.1}
\label{table:result_vvc_proposed_hevc}
\begin{tabular}{|c|c|c|} \hline
{Test Dataset}          &     Proposed solution        &    VVC (VTM-10.0)  \\ \hline \hline
{CTC (Overall)}         &        -10.21\%              &      -26.02\%      \\ \hline \hline
{Kodak (Overall)}       &        -11.63\%              &      -20.58\%      \\ \hline
\end{tabular}
\end{table}

\begin{table}[!t]
\centering
\caption{{Test: VVC (VTM-10.0), Anchor: HEVC (HM-16.20)}}
\renewcommand{\arraystretch}{1.1}
\label{table:result_HEVC_vs_VTM}
\begin{tabular}{|c|c|c|c|c|} \hline
{Test Dataset}           &    Y-BDR      &     U-BDR     &    V-BDR      &     CBDR    \\ \hline \hline
{CTC-Class A1}           &   -30.31\%    &   -31.74\%    &  -33.37\%     &  -30.63\%   \\ \hline
{CTC-Class A2}           &   -29.82\%    &   -22.33\%    &  -17.67\%     &  -28.42\%   \\ \hline
{CTC-Class B}            &   -23.29\%    &   -25.41\%    &  -29.39\%     &  -23.87\%   \\ \hline
{CTC-Class C}            &   -23.54\%    &   -19.73\%    &  -23.22\%     &  -23.24\%   \\ \hline
{CTC-Class E}            &   -26.52\%    &   -24.66\%    &  -25.18\%     &  -26.29\%   \\ \hline \hline
\bf{CTC (Overall)}       &   -26.14\%    &   -24.57\%    &  -26.03\%     &  -26.02\%   \\ \hline \hline
\bf{Kodak (Overall)}     &   -18.85\%    &   -37.01\%    &  -24.90\%	 &  -20.58\%   \\ \hline
\end{tabular}
\end{table}

{
\subsection{Comparison of encoder and decoder runtimes}
\label{subsec:compare_runtimes}
Table \ref{table:runtimes} presents the ratio of encoder and decoder runtimes of DLEC and VVC to that of HEVC estimated for different input video resolutions, where we specifically report:
\begin{align*}
    \mathsf{T}_{\text{DLEC}}^{E}=&\frac{\text{runtime of DLEC encoder}}{\text{runtime of HM encoder}}  \\[8pt]  
    \mathsf{T}_{\text{VVC}}^{E}=&\frac{\text{runtime of VTM-10.0 encoder}}{\text{runtime of HM encoder}}  \\[8pt]  
    \mathsf{T}_{\text{DLEC}}^{D}=&\frac{\text{runtime of DLEC decoder}}{\text{runtime of HM decoder}}  \\[8pt]  
    \mathsf{T}_{\text{VVC}}^{D}=&\frac{\text{runtime of VTM-10.0 decoder}}{\text{runtime of HM decoder}} 
\end{align*} 
All the runtime measures are obtained by running each codec on CTC sequences. In DLEC simulations, \emph{NVIDIA Tesla V100 (32GB)} GPUs are used, and each HM-16.20 and VTM-16.20 simulation is run on a single core of \emph{Intel Xeon Processor E5-1650 v3} CPU.
}

{Table \ref{table:runtimes} shows that both VTM and HEVC decoders require significantly less runtime than the proposed DLEC decoder. On the other hand, the DLEC encoder can provide about 2 to 30 times faster encoding than VTM depending on the resolution while HM being the fastest. It is important to note that the runtime measures of codecs can substantially differ depending on the software/hardware platform with their optimized implementations. Hence, the reported runtime results provide a general view on complexity by benchmarking against the reference HEVC and VVC codec implementations.}

\begin{table}[!t]
\centering
\caption{{Ratio of encoder/decoder runtimes of DLEC and VVC (VTM-10.0) to the runtime of HEVC (HM-16.20) estimated for  different video resolutions.}}
\renewcommand{\arraystretch}{1.3}
\label{table:runtimes}
\begin{tabular}{|c||c|c||c|c|} \hline
\multirow{ 2}{*}{Resolution} & \multicolumn{2}{c||}{Encoder}   
             & \multicolumn{2}{c|}{Decoder}   \\ \cline{2-5}
             & $\mathsf{T}_{\text{DLEC}}^{E}$  
             & $\mathsf{T}_{\text{VVC}}^{E}$   
             & $\mathsf{T}_{\text{DLEC}}^{D}$   
             & $\mathsf{T}_{\text{VVC}}^{D}$   \\ \hline \hline
3840$\times$2160  & 	8.1   &  31.4 &  709.9 &  	2.0   \\ \hline
1920$\times$1080  &	    5.3	  &  36.6 &  539.3 &  	1.9   \\ \hline
1280$\times$720   &	   10.3   &  20.2 &  991.4 &  	1.7   \\ \hline
832$\times$480    &	    2.3	  &  38.7 &  190.6 &  	2.1   \\ \hline
\end{tabular}
\end{table}


\section{Conclusions}
\label{sec:conclusion}
In this work, we investigated several transform network solutions to effectively support chroma subsampled YUV 4:2:0 format in VAE-oriented DLEC approaches. The performance of introduced solutions are tested on an evaluation framework based on the common test conditions (CTC) used in VVC standardization activities. Since HEVC and VVC are primarily designed for YUV 4:2:0 format, the compression results presented in this paper provide a more fair comparison against the state-of-the-art standards than the results reported in existing studies focusing only on coding RGB sources. 

The inspection of our comprehensive experimental results have led us to the following conclusions:
\begin{itemize}
    \item The proposed transform architecture in Fig.\ref{fig:arc3_allprelu} achieves about a significant 7\% average coding gains over naive input-output channel alignment solutions introduced in Section \ref{subsec:naive_alignment}. Therefore, the proposed new transform network architecture is essential for efficient coding of YUV 4:2:0 sources. 
    \item The proposed transform network solution achieves about 10\% average coding gains over HEVC, while it is 16\% less efficient than VVC. The 16\% performance gap against VVC can be further diminished by building more sophisticated hyperprior network architectures and latent variable modeling, which is considered as part of the future work.  
    \item The state-of-the-art GDN operators used in many DLEC approaches can be replaced with PReLU activations by including a pair of $1\times1$ convolutions as in Fig.\ref{fig:arc3_allprelu}, so that the resulting transform network design (i) is less complex (see Table \ref{table:complexity}) and (ii) provides about a substantial 3.8\% average coding benefit over only using GDNs (see Tables \ref{table:result_Sep_vs_joint_GDN} and \ref{table:result_Sep_vs_joint_pReLU_allRemoved}) {on CTC dataset \cite{Bossen:18:ctc} in YUV 4:2:0 format}.
    \item Since our solutions directly operate in YUV 4:2:0 domain, the fidelity in Y, U and V components can be adjusted by training network parameters using loss functions with different component-wise weighted distortion terms such as in (\ref{eqn:loss_initial}) and (\ref{eqn:loss_reweighted}).
    \item Based on the results on CTC video dataset, DLEC approaches tend to perform better at low-bitrate regimes (e.g., bitrates less than 5000 kbps).
    \end{itemize}

\bibliographystyle{IEEEtran}
\bibliography{refs}

\end{document}